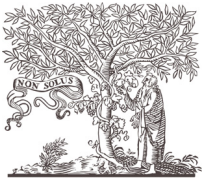

Computer Law
&
Security Review

# The flaws of policies requiring human oversight of government algorithms

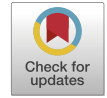

*Ben Green*

*University of Michigan, Ann Arbor, MI, USA*

## ARTICLE INFO

## ABSTRACT

As algorithms become an influential component of government decision-making around the world, policymakers have debated how governments can attain the benefits of algorithms while preventing the harms of algorithms. One mechanism that has become a centerpiece of global efforts to regulate government algorithms is to require human oversight of algorithmic decisions. Despite the widespread turn to human oversight, these policies rest on an uninterrogated assumption: that people are able to effectively oversee algorithmic decision-making. In this article, I survey 41 policies that prescribe human oversight of government algorithms and find that they suffer from two significant flaws. First, evidence suggests that people are unable to perform the desired oversight functions. Second, as a result of the first flaw, human oversight policies legitimize government uses of faulty and controversial algorithms without addressing the fundamental issues with these tools. Thus, rather than protect against the potential harms of algorithmic decision-making in government, human oversight policies provide a false sense of security in adopting algorithms and enable vendors and agencies to shirk accountability for algorithmic harms. In light of these flaws, I propose a shift from human oversight to institutional oversight as the central mechanism for regulating government algorithms. This institutional approach operates in two stages. First, agencies must justify that it is appropriate to incorporate an algorithm into decision-making and that any proposed forms of human oversight are supported by empirical evidence. Second, these justifications must receive democratic review and approval before the agency can adopt the algorithm.

## 1. Introduction

In recent years, governments across the world have turned to automated decision-making systems, often described as algorithms, to make or inform consequential decisions (Calo & Citron, 2021; Eubanks, 2018; Green, 2019; Henley & Booth, 2020). These developments have raised significant debate about when and how governments should adopt algorithms. On the one hand, algorithms bring the promise of making decisions more accurately, fairly, and consistently than public servants (Kleinberg et al., 2018; Kleinberg et al., 2015). On the other hand, the use of algorithms by governments has been a source of numerous injustices (Angwin et al., 2016; Calo & Citron, 2021; Eubanks, 2018; Green, 2019; Henley & Booth, 2020). The algorithms used in practice tend to be rife with errors and biases, leading to decisions that are based on incorrect information and that exacerbate inequities. Furthermore, making decisions via the rigid, rule-based logic of algorithms violates the principle that government decisions should respond to the circumstances of individual people.

In the face of these competing hopes and fears about algorithmic decision-making, policymakers have explored regulatory approaches that could enable governments to attain the benefits of algorithms while avoiding the risks of algorithms. An emerging centerpiece of this global regulatory effort is to require human oversight of the decisions rendered by algorithms. Human oversight policies enable governments to use algorithms—but only if a human has some form of oversight or control over the final decision.[1] In other words, algorithms may assist human decision-makers but may not make final judgments on their own. These policies are intended to ensure that a human plays a role of quality control, protecting against mistaken or biased algorithmic predictions. These policies aim to protect human rights and dignity by keeping a "human in the loop" of automated decision-making (Jones, 2017; Wagner, 2019). In theory, adopting algorithms while ensuring human oversight could enable governments to obtain the best of both worlds: the accuracy, objectivity, and consistency of algorithmic decision-making paired with the individualized and contextual discretion of human decision-making.

Recent legislation assumes that human oversight protects against the harms of government algorithms. For instance, in its proposed Artificial Intelligence Act, the European Commission asserted that human oversight (along with other mechanisms) is "strictly necessary to mitigate the risks to fundamental rights and safety posed by AI" (European Commission, 2021). Following this logic, many policies position human oversight as a distinguishing factor that makes government use of algorithms permissible. The European Union's General Data Protection Regulation (GDPR) restricts significant decisions "based solely on automated processing" (European Parliament & Council of the European Union, 2016b). The Government of Canada requires federal agencies using high-risk AI systems to ensure that there is human intervention during the decision-making process and that a human makes the final decisions (Government of Canada, 2021). Washington State allows state and local government agencies to use facial recognition in certain instances, but only if high-impact decisions "are subject to meaningful human review" (Washington State Legislature, 2020).

Despite the emphasis that legislators have placed on human oversight as a mechanism to mitigate the risks of government algorithms, the functional quality of these policies has not been thoroughly interrogated. Policymakers calling for human oversight invoke values such as human rights and dignity as a motivation for these policies, but rarely reference empirical evidence demonstrating that human oversight actually advances those values.[2] In fact, when policies and policy guidance do reference empirical evidence about human-algorithm interactions, they usually express reservations about the limits of human oversight, particularly related to people over-relying on algorithmic advice (Engstrom et al., 2020; European Commission, 2021; UK Information Commissioner's Office, 2020).

This lack of empirical grounding raises an important question: does human oversight provide reliable protection against algorithmic harms? Although inserting a "human in the loop" may appear to satisfy legal and philosophical principles, research into sociotechnical systems demonstrates that people and technologies often do not interact as expected (Suchman et al., 1999). Hybrid systems that require collaboration between humans and automated technologies are notoriously difficult to design, implement, and regulate effectively (Bainbridge, 1983; Gray & Suri, 2019; Jones, 2015; Pasquale, 2020; Perrow, 1999). Thus, given that human oversight is being enacted into policies across the world as a central safeguard against the risks of government algorithms, it is vital to ensure that human oversight actually provides the desired protections. If people cannot oversee algorithms as intended, human oversight policies would have the perverse effect of alleviating scrutiny of government algorithms without actually addressing the underlying concerns.

This article interrogates the efficacy and impacts of human oversight policies. It proceeds in four parts. The first two parts lay out the context of my analysis. Section 2 provides background on the tensions and challenges raised by the use of algorithms in government decision-making. Section 3 describes the current landscape of human oversight policies. I survey 41 policy documents from across the world that provide some form of official mandate or guidance regarding human oversight of public sector algorithms. I find that these policies prescribe three approaches to human oversight.

Section 4 evaluates the three forms of human oversight described in Section 3. Drawing on empirical evidence about how people interact with algorithms, I find that human oversight policies suffer from two significant flaws. First, human oversight policies are not supported by empirical evidence: the vast majority of research suggests that people cannot reliably perform any of the desired oversight functions. This first flaw leads to a second flaw: human oversight policies legitimize the use of flawed and unaccountable algorithms in government. Thus, rather than protect against the potential harms of algorithmic decision-making in government, human oversight policies create a regulatory loophole: it provides a false sense of security in adopting algorithms and enables vendors and agencies to foist accountability for algorithmic harms onto lower-level human operators.

Section 5 suggests how to adapt regulation of government algorithms in light of the two flaws to human oversight policies described in Section 4. It is clear that policymakers must stop relying on human oversight as a remedy for the potential harms of algorithms. However, the correct response is not to simply abandon human oversight, leaving governments to depend on autonomous algorithmic judgments. Nor should regulators prohibit governments from ever using algorithms. Instead, legislators must develop alternative governance approaches that more rigorously address the concerns which motivate the (misguided) turn to human oversight.

---

[1] Throughout this paper, "human oversight" refers to human judgment at the moment an algorithm renders a specific prediction or decision. An example of this form of human oversight involves a judge deciding whether to follow a pretrial risk assessment algorithm's recommendation to release a criminal defendant before trial. Human oversight is therefore distinct from more democratic and institutional approaches to having people govern algorithms, such as people's councils (McQuillan, 2018) and public task forces (Richardson, 2019).

[2] This lack of evidence is particularly striking given that governments often justify using algorithms as an "evidence-based" practice (New Jersey Courts, 2017; Starr, 2014).

I propose a shift from human oversight to institutional oversight for regulating government algorithms. This involves a two-stage process. First, rather than assume that human oversight can address fundamental concerns about algorithmic decision-making, agencies must provide written justification of its decision to adopt an algorithm in high-stakes decisions. If an algorithm violates fundamental rights, is badly suited to a decision-making process, or is untrustworthy, then governments should not use the algorithm, even with human oversight. As part of this justification, agencies should provide evidence that any proposed forms of human oversight are supported by empirical evidence. If there is not sufficient evidence demonstrating that human oversight is effective and that the algorithm improves human decision-making, then governments should not incorporate the algorithm into human decision-making processes. Second, agencies must make these written justifications publicly available. Agencies are not allowed to use the algorithm until its report receives public review and approval.

Compared to the status quo of blanket rules that enable governments to use algorithms as long as a human provides oversight, this institutional oversight approach will help to prevent human oversight from operating as a superficial salve for the injustices associated with algorithmic decision-making.

## 2. Discretion, algorithms, and decision-making in government

Human oversight policies arise from the conflicting hopes and fears about algorithmic decision-making in government. Many of the most consequential and controversial government uses of algorithms take place in street-level bureaucracies such as courts, police departments, schools, welfare agencies, and social service agencies. Within street-level bureaucracies, street-level bureaucrats—such as judges, police officers, teachers, and social workers—make consequential decisions about how to apply public policy to specific individuals (Lipsky, 2010). Examples of algorithms in street-level bureaucracies include judges using risk assessments to inform pretrial and sentencing decisions (Angwin et al., 2016; Wisconsin Supreme Court, 2016), child services workers using predictive models to inform which families to investigate for child neglect and abuse (Eubanks, 2018), and welfare agencies using algorithms to determine eligibility for benefits (Allhutter et al., 2020; Calo & Citron, 2021; Charette, 2018; Henley & Booth, 2020). Policymakers call most strongly for human oversight of algorithms in high-stakes decisions such as these (European Commission, 2021; European Parliament & Council of the European Union, 2016b; Government of Canada, 2021).

More than any other government setting, street-level bureaucracies are caught in a dilemma between rule-based and standard-based decision-making. While rules involve clear definitions and consequences, facilitating consistency and predictability, standards permit discretion, facilitating flexibility and sensitivity to the circumstances of specific cases (Solum, 2009). Street-level bureaucrats thus face an "essential paradox" (Lipsky, 2010). On the one hand, they must strive to treat everyone equally by following prespecified rules. On the other hand, they must also strive to be responsive to individual cases by exercising discretion. Adherence to either of these goals necessarily conflicts with the other (Lipsky, 2010). Furthermore, there is "no 'objective' solution" regarding the appropriate balance between rules and discretion (Wilson, 2000).

The extent to which judges and other street-level bureaucrats should be granted discretion over individual cases is thus a source of strenuous and recurring debate (Christie, 1986; Lipsky, 2010). On the one hand, publics and policymakers often express a strong desire to curb discretion in the interest of objectivity and consistency (Christie, 1986; Lipsky, 2010; Wilson, 2000). Allowing human decision-makers to exercise discretion means relying on the judgment and morality of those individuals, a prospect that causes significant unease (Zacka, 2017). Such discretion raises the specter of biased or arbitrary treatment by unelected agents of the state (Zacka, 2017). The United States is particularly prone to using rules to limit bureaucratic discretion (Wilson, 2000).

On the other hand, there are many reasons to desire discretion in street-level bureaucracies. Given that public policies are laden with ambiguous and conflicting goals, discretion enables street-level bureaucrats to make decisions based on the particular context at hand (Zacka, 2017). Restricting discretion would entail imposing a rigid, formal logic on the unavoidably ambiguous, uncertain, and unpredictable situations that street-level bureaucrats encounter, preventing them from adapting government decisions to complex or novel situations (Lipsky, 2010; Zacka, 2017). Furthermore, requiring judges, police officers, social workers, and other street-level bureaucrats to follow prespecified rules would violate the deeply held normative commitment that decisions should be responsive to individuals' conditions and needs (Lipsky, 2010). Discretion is therefore desirable—both normatively and practically—as it allows decision-makers to strike an appropriate balance between a variety of important values and objectives in light of each individual's particular circumstances (Lipsky, 2010; Zacka, 2017).

These competing demands on street-level bureaucracies lead to significant debate about the proper roles for algorithms in government. The desire to reduce individual discretion and promote consistency present strong motivations for government agencies to use algorithms. Algorithms bring a promise of accuracy, objectivity, and consistency that is attractive to both policymakers and publics. Evidence suggests that algorithms make policy-relevant predictions more accurately, fairly, and consistently than human servants (Kleinberg et al., 2018; Kleinberg et al., 2015). Thus, in addition to goals such as cutting costs and enhancing efficiency, governments adopt algorithms to attain greater accuracy when making predictions, replace biased human decisions with "objective" automated ones, and promote more consistent decision-making (Calo & Citron, 2021; Engstrom et al., 2020; Green, 2019; New Jersey Courts, 2017).[3]

---

[3] These efforts in government also reflect a broader political economic trend toward replacing or undercutting workers with AI systems (Gray & Suri, 2019; Pasquale, 2020). Across domains, however, AI perennially relies on human assistance (Gray & Suri, 2019)

However, despite these desires for algorithmic accuracy, objectivity, and consistency, the specter of governments making high-stakes decisions with algorithms raises two significant concerns. First, evidence suggests that these algorithms are neither as accurate nor fair as hoped. Algorithms used in settings such as the criminal justice system (Angwin et al., 2016), education (Kolkman, 2020), policing (Fussey & Murray, 2020; Green, 2019), and welfare (Calo & Citron, 2021; Charette, 2018) have suffered from notably high error rates. Many of these and other algorithms used by governments are biased against women, minorities, and low-income individuals (Angwin et al., 2016; Buolamwini & Gebru, 2018; Kolkman, 2020; Richardson et al., 2019).

Second, making decisions via the rigid, rule-based logic of algorithms violates the principle that government decisions should respond to individual circumstances. Street-level bureaucracies and administrative agencies derive their legitimacy from their expertise, discretion, and flexibility to shape policy based on complex and shifting circumstances (Calo & Citron, 2021). Yet unlike human decision-makers, algorithms adhere to predetermined decision rules and cannot reflexively adapt to novel or marginal circumstances (Alkhatib & Bernstein, 2019). Scholars have thus raised concerns that automation and algorithms significantly reduce expertise and discretion in street-level bureaucracies and administrative agencies (Bovens & Zouridis, 2002; Buffat, 2015; Calo & Citron, 2021). Although it may be inadvisable to provide individuals with a general right to have decisions about them be made by humans (Huq, 2020), individual justice requires human judgment for decisions that involve ethical and contextual analyses (Binns, 2020). This philosophically grounded notion is matched by human perceptions that decisions made by algorithms are less trustworthy and fair than those made by humans and that being evaluated by algorithms is dehumanizing (Binns et al., 2018; Lee, 2018). Automated decision-making also threatens due process, as the subjects of decisions often do not receive notice or the ability to meaningfully inspect and challenge decisions (Citron, 2008).

Thus, although algorithms promise certain benefits, algorithmic decision-making raises significant concerns about the unreliability of algorithms and the lack of contextual human discretion in decision-making. Government efforts to adopt algorithms are therefore often met with public skepticism and resistance, as communities reject the prospect of inhuman and inflexible automated systems shaping consequential decisions about their lives. For instance, protests against such tools in recent years have included signs saying "Families over Algorithms" (Scharfenberg, 2018) and chants of "Fuck the algorithm" (Kolkman, 2020).

These concerns motivate the turn to human oversight. These policies appear to present a responsible compromise between the promises and perils of algorithmic decision-making in government. By requiring that a human is kept in the decision-making loop, human oversight policies attempt to obtain the benefits of algorithms (accuracy, objectivity, and consistency) while protecting against the risks of algorithms

(errors, biases, and a lack of individualized judgment) and maintaining the attributes that are essential to street-level bureaucracies and administrative agencies (discretion, expertise, and due process). The rest of this article describes and evaluates these policies.

## 3. Survey of human oversight policies

This section summarizes how legislation and policy guidance describe the appropriate role for human oversight of government algorithms. To conduct this survey, I collected policy documents that fall into three categories: 1) proposed or passed legislation; 2) policy guidance by government or government-appointed bodies; and 3) manuals, policies, and court cases related to two notably controversial and high-stakes risk assessment tools in the US (risk assessments used in criminal justice settings and the Allegheny County Family Screening Tool). I discovered documents by searching for recently passed or proposed legislation related to AI and privacy, reviewing academic literature and news stories regarding AI regulation, and searching for documentation from the vendors and managers of criminal justice risk assessments and the Allegheny County Family Screening Tool. I reviewed each document to determine whether it discusses human use or oversight of algorithms. This process yielded 41 policy documents that provide some form of official mandate or guidance regarding human oversight of public sector algorithms.[4] These 41 documents are listed in the Appendix (and are all cited at least once in this section).[5]

I analyzed these policy documents to determine how they describe the proper role for humans in government decision-making processes aided by algorithms. I used inductive coding, looking particularly for what each document emphasizes as the central principle for human oversight. I also looked for any specific mechanisms that these documents propose regarding how to facilitate the desired form of human involvement and oversight in decision-making.

This coding process revealed that policies take three approaches to human oversight. Each approach rests on a central word or phrase that links the documents to a common principle for human oversight.[6] I will describe these approaches in order from least to most stringent (in terms of the requirements placed on human oversight). I present quotes from the

---

and more just outcomes arise when AI is used to complement rather than replace professionals (Pasquale, 2020).

[4] Although many of these documents are specifically focused on government uses of algorithms, others provide broader guidance that encompasses—but does not exclusively consider—government settings. As such, much of the analysis that follows also applies to human oversight in non-government settings.

[5] Out of these 41 documents, 23 are proposed or passed legislation, eight provide policy guidance, and ten are related to the two controversial risk assessment tools.

[6] The policy documents following each approach share notably consistent words and phrases. This pattern simplified the coding process. It also follows broader trends in technology policy development. Recent privacy laws have followed the model of prior, high-profile privacy laws such as Europe's General Data Protection Regulation (GDPR) and the California Consumer Privacy Act (CCPA) (Chander et al., 2021; Schwartz, 2019). A similar pattern has been observed in AI ethics, as recent statements of principles share many common elements (Fjeld et al., 2020; Jobin et al., 2019).

applicable policy documents in order to demonstrate the coherence of each approach.

### 3.1. Restricting "solely" automated decisions

The first approach to human oversight is to directly prohibit or restrict decisions that are made through "solely" automated means. Twenty of the 41 policy documents take this approach to oversight. All of them are proposed or passed legislation.

The European Union's General Data Protection Regulation (GDPR) is the most notable example of a law that restricts solely automated decisions. Article 22 of the GDPR mandates (with several exceptions) that "data subject[s] shall have the right not to be subject to a decision based solely on automated processing, including profiling, which produces legal effects concerning him or her or similarly significantly affects him or her" (European Parliament & Council of the European Union, 2016b). Eight of the EU member states (plus the United Kingdom) specifically address solely automated decision-making in their national GDPR implementations.[7] The EU Law Enforcement Directive applies similar restrictions as GDPR Article 22 to automated processing by criminal law enforcement agencies (European Parliament & Council of the European Union, 2016a).

Since the passage of the GDPR, many jurisdictions outside the EU have proposed or passed laws that include GDPR-style restrictions on solely automated decision-making. Policies in Argentina (Republic of Argentina, 2018), Mauritius (Parliament of Mauritius, 2017), Kenya (Republic of Kenya, 2019), and South Africa[8] (Republic of South Africa, 2013) provide default prohibitions on solely automated decision-making, much like the GDPR. Policies in Bahrain (Kingdom of Bahrain, 2018), Brazil (National Congress of Brazil, 2019), Québec (National Assembly of Québec, 2020), Uganda (The Republic of Uganda, 2019), and the United States (Brown, 2020) do not prohibit any instances of solely automated decision-making, but require protections for the subjects of solely automated decisions.

All of the policies discussed in this section require that the subjects of any solely automated decisions are granted rights and protections. One of the most emphasized and common safeguards is the right for subjects of solely automated decisions to obtain post hoc human intervention. In other words, after someone has been subject to a solely automated decision, they can request that a human inspect and consider altering that decision. Twelve of the 19 policies discussed in this section explicitly incorporate a right to human intervention.[9]

### 3.2. Emphasizing human discretion

The second approach to human oversight reflects a corollary to the first approach. In these documents, policymakers, algorithm vendors, and courts emphasize that decisions must involve human discretion. Fourteen of the 41 policy documents take this approach to oversight. One is passed legislation, three provide policy guidance, and ten relate to criminal justice and child welfare risk assessments.

Several policy documents state that human oversight and discretion are essential for protecting values such as human rights. The Canadian Directive on Automated Decision-Making requires that federal agency decisions likely to have "high" or "very high" social impacts "cannot be made without having specific human intervention points during the decision-making process; and the final decision must be made by a human" (Government of Canada, 2021). A discussion paper by the Australian Human Rights Commission proposes that algorithmic decisions "must be [...] subject to appropriate human oversight and intervention" (Australian Human Rights Commission, 2019). Multiple reports by the New Zealand Government similarly emphasize the need to "retain" and "preserve" human oversight as core priorities for how governments manage algorithms (Statistics New Zealand, 2018, 2020).

All ten of the documents related to controversial risk assessment algorithms in the United States emphasize the importance of human discretion. In the face of significant public scrutiny, these documents present human discretion as a safety valve that limits the influence of algorithms and mitigates the potential harms of mistaken or biased predictions. These documents attempt to enable these protections by granting decision-makers the discretion to override algorithmic judgments.

The Allegheny County Department of Human Services, which oversees the Allegheny Country Family Screening Tool (AFST),[10] presents human oversight as an essential safeguard for decision-making. In response to concerns about the algorithm fully determining outcomes, the Department stresses that "[s]creening decisions are not in any way 'dictated' by the AFST," as "supervisors have full discretion over call screening decisions" (Allegheny County Department of Human Services, 2019a). The Department further justifies the AFST due to the use of human discretion, claiming that the tool produces "few, if any, unintended adverse effects given workers' will-ingness to use their own discretion in the screening decision" (Allegheny County Department of Human Services, 2019b).

---

[7] A recent article surveys how each EU member state implements Article 22, along with translations of the relevant sections from the laws that are not otherwise available in English (Malgieri, 2019). Austria (Austrian Parliament, 2018), Belgium (Belgian Federal Parliament, 2018), France (French Parliament, 2018), and Hungary (Hungarian Parliament, 2018) expand the scope of restrictions on solely automated decision-making. France (French Parliament, 2018), Germany (Bundestag, 2019), and the Netherlands (Dutch Parliament, 2018) distinguish the requirements for particular public and private sector settings. Ireland (Houses of the Oireachtas, 2018) and the United Kingdom (UK Parliament, 2018) describe detailed procedures that must accompany any permitted solely automated decisions (The UK's implementation of the GDPR went into effect in 2018, before the UK left the EU, and remains in effect). Slovenia (Slovenian Parliament, 2020) calls for ex ante impact assessments.

[8] South Africa passed the Protection of Personal Information Act in 2013, following the release of the draft GDPR. Most of the provisions (including the section related to automated decision-making) went into effect in 2020.

[9] The seven exceptions are Argentina, Austria, France, Mauritius, Québec, Slovenia, and South Africa.

[10] The AFST is a risk assessment that predicts the likelihood that children will be removed from their home due to child neglect and abuse in the next two years. These predictions are presented to child welfare workers to inform their decisions about which cases to investigate (De-Arteaga et al., 2020; Eubanks, 2018).

Documents related to criminal justice risk assessments place a similar emphasis on discretion. Northpointe, which developed the COMPAS risk assessment,[11] acknowledges that the algorithm can make mistakes and writes that "staff should be encouraged to use their professional judgment and override the computed risk as appropriate" (Northpointe, 2015). Arnold Ventures, which developed the Public Safety Assessment (PSA), strongly emphasizes that "[j]udges are *not required* to follow the PSA" (Arnold Ventures, 2019). The New Jersey Courts (which adopted the PSA in 2017) writes that the PSA "do[es] not replace judicial discretion" (New Jersey Courts, 2017). Other organizations that create and oversee criminal justice risk assessments similarly highlight the importance of professional judgment and the ability of staff to override a risk assessment's recommendations (Andrews & Bonta, 2001; Steinhart, 2006; Wisconsin Department of Corrections, 2018).

Human discretion also played a central role in two court cases that supported the use of risk assessments in criminal sentencing. In *Malenchik v. State of Indiana* and *State of Wisconsin v. Loomis*, two state supreme courts considered whether it was appropriate for risk assessments to inform sentencing decisions. Both courts argued that as long as judges have discretion regarding how to incorporate algorithmic advice into sentences, then it is acceptable to use risk assessments to inform criminal sentencing. Noting that risk scores are "neither […] intended nor recommended" to replace individualized sentencing decisions, the Indiana Supreme Court stated that it "defer[s] to the sound discernment and discretion of trial judges to give the tools proper consideration and appropriate weight" (Indiana Supreme Court, 2010). Similarly, acknowledging that COMPAS is imperfect, the Wisconsin Supreme Court stated that "courts [should] exercise discretion when assessing a COMPAS risk score with respect to each individual defendant" (Wisconsin Supreme Court, 2016).

### 3.3. Requiring "meaningful" human input

The third approach to human oversight represents an extension of the first and second approaches. Rather than simply calling for human involvement in decision-making, documents taking this third approach recognize that some forms of human involvement can be superficial or inadequate. These policies therefore emphasize that human input and oversight must be "meaningful." Seven of the 41 reviewed documents take this approach to oversight. Two are proposed or passed legislation and five provide policy guidance.

Efforts to promote "meaningful" human input are intended to avoid the pitfalls of restrictions on "solely" automated decisions. As Section 4.1.1 will discuss in more detail, the narrow scope of such restrictions makes it possible for institutions to circumvent regulatory obligations by inserting superficial forms of human involvement into the decision-making process (Veale & Edwards, 2018; Wagner, 2019).

Two influential European bodies have emphasized meaningful human input in direct reference to the GDPR. In its guidance related to the GDPR, the Article 29 Data Protection Working Party asserts that "[t]o qualify as human involvement, the controller must ensure that any oversight of the decision is meaningful, rather than just a token gesture" (Article 29 Data Protection Working Party, 2018). The UK Information Commissioner's Office stresses that "human input needs to be **meaningful**," clarifying that "a decision does not fall outside the scope of [GDPR] Article 22 just because a human has 'rubberstamped' it" (UK Information Commissioner's Office, 2020).

Other policies and policy guidance also emphasize meaningful human oversight of government algorithms. A Washington State law requires that any government decisions that involve facial recognition "are subject to meaningful human review" (Washington State Legislature, 2020). The European Commission's High-Level Expert Group on AI lists "human agency and oversight" as the first of "seven key requirements for Trustworthy AI" and emphasizes the importance of "meaningful opportunity for human choice" (High-Level Expert Group on AI, 2019). The European Commission's proposal for an Artificial Intelligence Act and a report commissioned by the Administrative Conference of the United States similarly stress the need to avoid simplistic and superficial forms of human oversight (Engstrom et al., 2020; European Commission, 2021).

Although none of these policy documents provide precise or detailed definitions of "meaningful" human involvement, they suggest three components that are central to meaningful human oversight. First, human decision-makers must be able to disagree with the algorithm's recommendations. In other words, for human oversight to be meaningful, human reviewers must have the competence and authority to override algorithmic decisions (Article 29 Data Protection Working Party, 2018; European Commission, 2021; High-Level Expert Group on AI, 2019; UK Information Commissioner's Office, 2020; Washington State Legislature, 2020).[12]

Second, human overseers must understand how the algorithm operates and makes decisions. Several policy guidance documents suggest explanations of algorithmic decisions as a mechanism to facilitate human understanding (Engstrom et al., 2020; High-Level Expert Group on AI, 2019; UK Information Commissioner's Office, 2020). Similarly, numerous policies and reports call for transparency into the functioning of algorithms so that human decision-makers can interpret the output of algorithmic systems (Engstrom et al., 2020; European Commission, 2021; High-Level Expert Group on AI, 2019; UK Information Commissioner's Office, 2020; Washington State Legislature, 2020).

Third, human decision-makers must not depend on algorithms and should instead thoroughly consider all of the information relevant to a given decision. Several documents raise concerns about people over-relying on algorithmic rec-ommendations even when granted the ability to make final decisions (Engstrom et al., 2020; European Commission, 2021; High-Level Expert Group on AI, 2020; UK Information Commissioner's Office, 2020). In addition, two European regulatory agencies stress that human reviewers must weigh algorithmic recommendations alongside all available information and all of the considerations relevant to a given decision (Article 29 Data Protection Working Party, 2018; UK Information Commissioner's Office, 2020).

---

[11] COMPAS (short for Correctional Offender Management Profiling for Alternative Sanctions) is a risk assessment that predicts the likelihood that criminal defendants will be arrested in the next two years. These predictions are presented to judges to inform their pretrial release and sentencing decisions (Angwin et al., 2016; Northpointe, 2015).

[12] In this respect, documents calling for meaningful human oversight align with the documents calling for human discretion described in Section 3.2.

## 4. Two flaws with human oversight policies

In this section, I evaluate the efficacy of human oversight policies and argue that they suffer from two flaws. First, drawing on recent empirical evidence about how people interact with algorithms in government and other settings, I consider whether people can provide the types of oversight that policies call for. Despite the hopes of policymakers, the vast majority of evidence suggests that people cannot provide the envisioned protections against algorithmic errors, biases, and inflexibility. Second, I consider the implications of these limits to human oversight. Ungrounded assumptions of effective human oversight promote a false sense of security in adopting algorithms and reduce the accountability that vendors and policymakers face for algorithmic harms.

### 4.1. Flaw 1: Human oversight policies are not supported by empirical evidence

The first flaw of human oversight policies is that they have little basis in empirical evidence. The vast majority of research suggests that people are unable to provide reliable oversight of algorithms. To show the empirical limits of human oversight policies, I will return to the three forms of human oversight introduced in Section 3 and describe why each is highly unlikely to provide the desired protections against algorithmic errors, biases, and inflexibility.

### 4.1.1. Restrictions on "solely" automated decisions provide superficial protection

Policies that restrict "solely" automated decisions have the clearest flaws. First, these policies provide protection for a limited number of cases. Public sector algorithms typically already operate with human involvement and with a human making the final decision, particularly in high-stakes settings such as criminal justice and child welfare. Policies restricting "solely" automated decisions therefore have no impact in many of the cases that generate public scrutiny and outcry.

Second, the narrow scope of "solely" automated decisions creates flimsy and easily avoidable protections. At least by the letter of these laws, any nominal form of human involvement is sufficient to avoid the protections placed on solely automated decisions. Provisions like the GDPR's Article 22 thus may create an incentive to introduce superficial human oversight of algorithms (i.e., "rubber stamping" automated deci-

sions) as a way to bypass regulations (Veale & Edwards, 2018; Wagner, 2019).[13]

In addition, the right to post hoc human intervention fails to provide robust protections against the harms of solely automated decisions. Procedurally, the right to human intervention puts the onus on individuals to request human review after they have already been harmed by a decision. Many people will have neither the means nor knowledge to take advantage of this right. Even when people do request human intervention, it can be slow and onerous to obtain remedies, allowing the harm of a flawed automated decision to manifest (Calo & Citron, 2021). Substantively, human intervention is unlikely to produce better decisions in most settings. This form of human intervention amounts to the ability of a human reviewer to override an automated decision. As Section 4.1.2 will describe, people are bad at evaluating the quality of algorithmic judgments, leading them to typically override algorithms in detrimental ways.

### 4.1.2. Human discretion does not improve outcomes

Even when human oversight moves beyond "rubber stamp" approaches, such that people have agency to use discretion and make final decisions, human oversight is unlikely to provide protections against the harms of algorithmic decision-making.

A long-standing body of research shows that, across a wide range of domains, automated decision-support systems tend to alter human decision-making in unexpected and harmful ways. Numerous studies have demonstrated that people (including experts) are susceptible to "automation bias"—i.e., people defer to automated systems, reducing the amount of independent scrutiny that they exhibit when making decisions (Parasuraman & Manzey, 2010; Skitka et al., 1999). Automation bias can involve omission errors—failing to take action because the automated system did not provide an alert—and commission errors—following the advice of an automated system even though it is incorrect and there is contradicting evidence (Parasuraman & Manzey, 2010; Skitka et al., 1999). Furthermore, automating certain parts of human tasks can make the remaining parts more difficult and cause human skills to deteriorate (Bainbridge, 1983). As a result, automated systems may simply lead to different types of errors rather than reducing overall errors as intended (Skitka et al., 1999). Automation can also create a diminished sense of control, responsibility, and moral agency among human operators (Berberian et al., 2012; Cummings, 2006).

More recent research demonstrates that similar issues arise when humans collaborate with predictive algorithms. Broadly speaking, people are bad at judging the quality of algorithmic outputs and determining whether and how to override those outputs. People struggle to evaluate the accuracy of algorithmic predictions (Goodwin & Fildes, 1999; Green & Chen, 2019a, 2019b; Lai & Tan, 2019; Springer et al., 2017), leading them to discount accurate algorithmic recommendations and to rely on inaccurate algorithmic recommenda-tions (Dietvorst et al., 2015; Goodwin & Fildes, 1999; Lim & O'Connor, 1995; Springer et al., 2017; Yeomans et al., 2017). This means that even though algorithmic advice can improve the accuracy of human predictions, people's judgments about when and how to diverge from algorithmic recommendations are typically incorrect (Green & Chen, 2019a, 2019b; Grgić-Hlača et al., 2019; Lai & Tan, 2019). People have also been shown to exhibit racial biases when incorporating algorithmic advice into their predictions (Green & Chen, 2019a, 2019b). Furthermore, although an evaluation of the Allegheny Family Screening Tool found that staff were able to override many algorithmic errors (De-Arteaga et al., 2020), other evidence shows that algorithmic errors reduce the quality of expert judgments (Kiani et al., 2020).

---

[13] This type of superficial human oversight would represent an example of "skeuomorphic humanity," providing the impression that a human is making decisions when that is not actually the case (Brennan-Marquez et al., 2019).

The use of algorithms in policing and the criminal justice system exemplifies the limits of human discretion in practice. Police have been shown to follow incorrect advice from algorithms, even when tasked with overseeing an algorithm and under no mandate to follow its advice. For instance, police in London "overwhelmingly overestimated the credibility" of a live facial recognition system, judging computer-generated matches to be correct at three times the actual rate of accuracy (Fussey & Murray, 2020). Such behavior led to the first known case of arrest due to faulty facial recognition in the United States, when the Detroit Police Department arrested a man due solely to a facial recognition match that was clearly incorrect (Hill, 2020).

In contrast, judges across the United States regularly deviate from algorithmic advice, but typically in detrimental ways. Evidence from several US jurisdictions shows that judges frequently override release recommendations in order to detain defendants, leading to inflated detention rates (Human Rights Watch, 2017; Sheriff's Justice Institute, 2016; Steinhart, 2006; Stevenson, 2018; Stevenson & Doleac, 2021). Furthermore, judges often make more punitive decisions regarding Black defendants than white defendants who have the same risk score, causing the introduction of risk assessments to exacerbate racial disparities in pretrial decisions (Albright, 2019; Cowgill, 2018; Stevenson & Doleac, 2021). Thus, rather than enable people to identify and correct algorithmic biases, human discretion can enable people to inject new forms of inconsistency and bias into decisions.

### 4.1.3.  Even "meaningful" human oversight does not improve outcomes

Finally, despite their rhetorical promises, policies mandating "meaningful" human oversight are unlikely to ensure protection against algorithmic harms. Such policies face two major issues. First is a definitional issue: none of these policies propose a definition of meaningful human oversight. Although policies agree that a human operator rubber stamping algorithmic decisions does not constitute meaningful oversight, they do not provide a standard for determining whether any particular form of human oversight is meaningful.[14] As a general matter, this could be appropriate: many policies rely on standards whose application depends on context (Lipsky, 2010; Solum, 2009; Zacka, 2017). However, for any definition of meaningful oversight to be desirable, at least some components of meaningful oversight must improve outcomes.

Thus, the first, definitional issue with meaningful human oversight leads to a second, functional issue: the three components described as central to meaningful human oversight are either unlikely to improve decision-making or are incredibly difficult to achieve. The first requirement of meaningful human oversight is that decision-makers must be allowed to override algorithmic recommendations. This is already the case in most high-stakes settings, such as the criminal justice system. Yet as described in Section 4.1.2, both laypeople and public servants tend to override algorithms in detrimental ways. Thus, although this proposed component of meaningful human oversight is satisfied in many instances, most evidence suggests that it does not improve outcomes.

The second requirement of meaningful human oversight is that human overseers must understand the algorithm's operations and outputs. Policymakers propose algorithmic explanations and transparency to facilitate this goal. As with the first requirement, despite the broad support for this idea, evidence suggests that algorithmic explanations and transparency do not actually improve human oversight. Studies have found that explanations do not improve people's ability to make use of algorithmic predictions (Bansal, Wu, et al., 2021; Green & Chen, 2019b). In fact, explanations can have the harmful effect of prompting people to place greater trust in algorithmic recommendations even when those recommendations are incorrect (Bansal, Wu, et al., 2021; Jacobs et al., 2021) or when the explanations have no basis in the algorithm's actual functioning (Lai & Tan, 2019). Algorithmic transparency similarly reduces people's ability to detect and correct model errors (Poursabzi-Sangdeh et al., 2021). Based on this evidence, the proposed remedies of explanations and transparency appear to hinder—rather than improve—people's ability to identify algorithmic mistakes and make effective use of algorithmic recommendations.

The third requirement of meaningful human oversight is that decision-makers must avoid relying on algorithms and instead consider all of the relevant information. Although this goal is appealing, evidence suggests that it is difficult (if not impossible) to achieve in practice. Studies have found that automation bias persists even after training and explicit instructions to verify an automated system (Parasuraman & Manzey, 2010). Furthermore, even when people do not rely entirely on automation, an algorithm can still significantly alter how people make decisions. Experimental studies suggest that risk assessments increase the weight that judges, law students, and laypeople place on risk relative to other considerations when making simulated pretrial and sentencing decisions (Green & Chen, 2021; Skeem et al., 2019; Starr, 2014). Without realizing it, decision-makers respond to algorithms by focusing more heavily on the considerations that algorithms highlight (Green & Chen, 2021). Thus, while it is desirable that decision-makers balance an algorithm's advice with other information and factors, evidence suggests that people typically defer to automated tools and increase their attention to the factors emphasized by algorithms.

---

[14]  In the related context of autonomous weapon systems, "meaningful human control" has gained widespread support as a governance principle, yet the "inherent imprecision" of this principle means that there is no consensus regarding what it actually entails (Crootof, 2016).

## 4.2. Flaw 2: Human oversight policies legitimize flawed and unaccountable algorithms in government

The second flaw of human oversight policies follows from the first: because human oversight does not protect against algorithmic harms, human oversight policies reduce scrutiny of government algorithms without reliably reducing the harms of these systems. This process has two dimensions. First, human oversight provisions provide policymakers and publics with a false sense of security that even flawed algorithms are safe to use in high-stakes arenas. Second, human oversight provisions shift accountability for algorithmic harms from agency leaders (who determine the structure of algorithmic systems) to frontline human operators (who are relatively powerless). Thus, in effect, human oversight policies create a loophole that allows agencies to adopt flawed algorithms and to shirk accountability for any harms that result.

### 4.2.1. The assumption of effective human oversight provides a false sense of security in adopting algorithms
Government algorithms raise significant risks about error-prone, biased, and inflexible algorithms making high-stakes policy decisions. These concerns call into question the desirability of adopting algorithms in government. In response, policymakers present human oversight as the salve that enables governments to obtain the benefits of algorithms without incurring the associated harms.[15] Were any of the proposed forms of human oversight effective, then perhaps this remedy would represent an effective compromise between the benefits and risks of government algorithms. Yet given the limits of human oversight, human oversight policies fail to mitigate the underlying concerns. Instead, these policies merely provide cover for fundamental concerns about the use of algorithms in government decision-making. In turn, human oversight policies justify the inappropriate integration of algorithms into government decision-making.

#### 4.2.1.1. Example 1: Human overrides of algorithmic decisions.
Two examples highlight how human oversight provides false comfort in the face of concerns that undermine the basis for algorithmic decision-making in government. The first example involves the assertion that people should regularly override algorithmic decisions. Risk assessment developers and managers point to human overrides as an important safeguard against imperfect predictions and as evidence that the algorithms are not replacing human judgment (Allegheny County Department of Human Services, 2019a; Northpointe, 2015; Wisconsin Department of Corrections, 2018). Similarly, multiple policy guidance documents calling for meaningful human oversight warn that, if humans agree with an algorithm too often, then decisions should be considered solely automated (Article 29 Data Protection Working Party, 2018; UK Information Commissioner's Office, 2020). In other words, for human oversight to be meaningful, decision-makers must routinely disagree with the automated system (Veale & Edwards, 2018; Wagner, 2019).

At first glance, these calls for overrides appear prudent. Policymakers are right to be concerned about the perils of solely automated decision-making. If a human decision-maker rarely overrides an algorithmic decision, then the decision-making process would be nearly solely automated, potentially violating due process and human dignity and subjecting people to mistaken and biased judgments. Allowing humans to disagree with algorithms seems to provide an avenue for injecting discretion and error-correction into decisions.

The problem, however, is that human overrides cannot actually remedy the concerns that motivate overrides. Policies calling for overrides therefore provide the appearance of quality control—legitimizing the use of flawed and controversial algorithms—but do not actually address the underlying issues.

Consider the two scenarios in which overrides of automated decisions appear particularly desirable. One reason to call for human overrides is a lack of trust in an algorithm to make accurate and fair decisions. In these cases, human overrides seem to provide quality control of algorithmic judgments. However, this remedy is unlikely to be effective: substantial evidence demonstrates that humans tend to override algorithms in detrimental rather than beneficial ways (Green & Chen, 2019a, 2019b; Grgić-Hlača et al., 2019; Lai & Tan, 2019).

A second reason to call for human overrides is that an algorithm fails to account for considerations that are essential to a given decision. For instance, pretrial risk assessments do not consider the full range of factors that judges must balance (Green & Chen, 2021). In these cases, human overrides seem to enable holistic judgments by incorporating considerations that the algorithm omits. However, this remedy is unlikely to be effective: people cannot reliably balance an algorithm's advice with other factors, as they often over-rely on automated advice (Parasuraman & Manzey, 2010; Skitka et al., 1999) and place greater weight on the factors that algorithms emphasize (Green & Chen, 2021; Skeem et al., 2019; Starr, 2014).

Thus, when policymakers call for human overrides, they deflect criticism of these tools but fail to mitigate the underlying concerns. Human oversight cannot address concerns about inaccurate, unfair, and myopic algorithms. More structural reforms are necessary. If an algorithm is so flawed that policymakers do not trust it to make decisions without a significant number of human overrides, then the appropriate remedy is to improve the algorithm. If the algorithm cannot be sufficiently improved, then the appropriate remedy is to stop using the algorithm. Similarly, if an algorithm ignores criteria that must be considered when making a given decision, then the appropriate remedy is to alter the algorithm so that it accounts for all relevant criteria. If this is not feasible, then the appropriate remedy is to stop using the algorithm.

#### 4.2.1.2. Example 2: Wisconsin v. Loomis.
The false sense of security provided by human oversight can also be seen in *State of*

---

[15] The "rubber stamp" loophole set up by prohibitions on "solely" automated decisions has proven relatively easy to identify (Veale & Edwards, 2018; Wagner, 2019), prompting some regulators developing legislation to drop calls for GDPR-style restrictions on solely automated decisions (Office of the Privacy Commissioner of Canada, 2020). However, the limits of the other two approaches to human oversight are more subtle and consequential, as these forms of oversight have intuitive appeal and are increasingly common.

*Wisconsin v. Loomis*. In this case, the Wisconsin Supreme Court ruled that courts could use the COMPAS risk assessment to inform sentencing as long as decisions involved judicial oversight and discretion (Wisconsin Supreme Court, 2016).[16] Although the Wisconsin Supreme Court was right to recognize the perils of relying on COMPAS to determine sentences, its reliance on human oversight to alleviate these concerns was misplaced.

Consider the two central worries that prompted the Wisconsin Supreme Court to call for human oversight. First, concerned about the errors and biases of risk assessments, the Court pointed to judges' discretion to override COMPAS. Most notably, the Court mandated that COMPAS be accompanied by a list of concerns that have been raised about the tool (Wisconsin Supreme Court, 2016). Although it seems reassuring to prompt judges to use discretion when considering risk assessments, doing so leaves judges with conflicting guidance: the Court hailed risk assessments for their ability to provide reliable and accurate predictions, yet also warned that risk assessments can be laden with errors and biases. Furthermore, discretion is unlikely to be an effective remedy, as judges often use their discretion to override risk assessments in punitive and racially biased ways, (Albright, 2019; Cowgill, 2018; Human Rights Watch, 2017; Sheriff's Justice Institute, 2016; Steinhart, 2006; Stevenson, 2018; Stevenson & Doleac, 2021).

Second, concerned about risk assessments violating due process, the Court attempted to limit the extent to which judges could rely on these tools. The Court asserted that judges may use risk assessments to inform—but not determine—certain aspects of sentences, defending the use of COMPAS in the case at hand because the risk assessment had no discernable impact on the sentencing decision. Even as it praised the ability of risk assessments to promote better outcomes, the Court affirmed Loomis' sentence on the grounds that the circuit court "would have imposed the same sentence regardless of whether it considered the COMPAS risk scores" (Wisconsin Supreme Court, 2016). Although it seems reassuring to assert that judges cannot rely on risk assessments, this reasoning again leaves judges with conflicting guidance. On the one hand, if risk assessments can be used only to inform outcomes that would have been reached independently, then there is no reason to use such tools at all. On the other hand, if risk assessments are to improve outcomes as the Court intends, then such tools must influence sentencing decisions. Furthermore, it is unlikely that judges can circumscribe their consideration of risk assessments as the Court envisions. Evidence suggests that judges and other people often defer to automated advice and change their decision-making processes due to algorithms, yet do not recognize that these behaviors are occurring (Green & Chen, 2021; Parasuraman & Manzey, 2010; Skeem et al., 2019; Starr, 2014).

Thus, by calling for judicial discretion, the Wisconsin Supreme Court alleviated its concerns about risk assessments but did not actually mitigate the harms associated with these tools. The foundational issue troubling the Court was the low quality of risk assessments and the conflict between risk assessments and due process. Judicial discretion cannot address these concerns. Rather than point to human oversight, the Court should have placed greater scrutiny on the algorithm's quality and on whether it is appropriate for an algorithm to alter a defendant's sentence. If the Court was not comfortable with the accuracy and fairness of risk assessments, then it should not have allowed these tools to be presented to judges until their quality improves. Similarly, if the Court was not comfortable with algorithms altering sentences, then it should not have allowed algorithms to be incorporated into sentencing adjudications at all. All told, if the Court would not allow the use of COMPAS without human oversight, then there is scant evidence supporting its decision to allow the use of COMPAS with human oversight.

### 4.2.2.  *Relying on human oversight diminishes responsibility and accountability for institutional decision-makers*

By appearing to address foundational concerns about government algorithms, human oversight policies shift responsibility for algorithmic systems from agency leaders and technology vendors to human operators. Human oversight policies position frontline human operators as the scapegoats for algorithmic harms, even though algorithmic errors and injustices are typically due to factors over which frontline human overseers have minimal agency, such as the system design and the political goals motivating implementation.[17] Indeed, even if human oversight were a reliable form of quality control, human oversight policies would still have the harmful effect of diminishing the accountability of agency leaders and vendors for their decisions to develop and implement algorithms.

The emphasis on human oversight as a protective mechanism allows governments and vendors to have it both ways: they can promote an algorithm by proclaiming how its capabilities exceed those of humans, while simultaneously defending the algorithm and those responsible for it from scrutiny by pointing to the security (supposedly) provided by human oversight. For instance, the Austrian Public Employment Service uses an algorithm that informs decisions about what forms of assistance to provide job seekers. The agency justifies this practice by hailing the algorithm's objectivity and precision. At the same time, in the face of public concern and acknowledged limitations, the agency legitimizes the algorithm by describing it as a mere "second opinion" that requires human review (Allhutter et al., 2020).

This dual rhetoric enables the actors responsible for developing and implementing an algorithm to attain goodwill for algorithmic benefits yet escape accountability for algorithmic harms. When something goes well, governments and vendors can praise the algorithm (as well as their own wisdom in adopting the algorithm). When something goes wrong, gov-ernments and vendors can blame and punish the individuals operating the system.

---

[16] The Indiana Supreme Court followed similar reasoning when justifying risk assessment tools in *Malenchik v. State of Indiana* (Indiana Supreme Court, 2010).

---

[17] This phenomenon matches the long-standing pattern of human operators being blamed for breakdowns in technical systems, even though the harms of these systems are typically structured by the decisions of more powerful institutional actors (Elish, 2019; Perrow, 1999). In similar manner, blaming car crashes on human error obscures the role of automobile manufacturers and traffic engineers in creating dangerous driving conditions (Zipper, 2021).

A notable instance of this convenient finger-pointing occurred in the aftermath of a Black man in Detroit being wrongfully arrested following an incorrect match by a Detroit Police Department (DPD) facial recognition system (Hill, 2020). Representatives from each of the three technology companies that produced the system immediately blamed the mistaken arrest on human operators following an inappropriate investigation process (Hill, 2020). Appearing the following year on the national news program *60 Minutes*, the Detroit police chief similarly blamed "[s]loppy, sloppy investigative work" for the incident (CBS News, 2021). Noting that the detective and commanding officer have since been disciplined, the Detroit Police Chief added, "it wasn't facial recognition that failed. What failed was a horrible investigation" (CBS News, 2021).

Although the operators surely could have followed a more thorough investigative process in this particular case, placing blame on the operators obscures the role of other actors—particularly the technology vendors and police chief—responsible for the system-level decisions that led to this arrest. It is DPD leadership and the technology vendors who chose to implement a shoddily-tested investigative technology known to have low accuracy on Black faces (Buolamwini & Gebru, 2018; Hill, 2020) in the US city with the largest share of Black residents, against the opposition of many (Campbell, 2019). In fact, the Detroit Police Chief himself admitted that the DPD's facial recognition system is incorrect 96% percent of the time, and DPD data demonstrates that the system is used almost exclusively to investigate Black suspects (Koebler, 2020).

Thus, although the human operators were the most proximate to the wrongful arrest, the police chief and vendors are more substantively responsible for the incident. They are the ones who should be held accountable. No form of human oversight could make it appropriate for DPD to use a facial recognition system that violates civil liberties, is incorrect in the vast majority of cases, and is used to surveil Detroit's Black population. Instead, these harms can be remedied only by banning police facial recognition altogether (Hartzog & Selinger, 2018; Stark, 2019), as several jurisdictions across the United States have recently done (Hill, 2021) and many civil society organizations have called for (Amnesty International, 2021; European Digital Rights, 2021a; Fight For The Future, 2021).

## 5. From human oversight to institutional oversight

This study has shown that policies mandating human oversight for government algorithms are flawed in two ways. First, human oversight is unable to provide the desired protections. Second, human oversight policies legitimize flawed and unaccountable algorithms in government without remedying the issues with these tools. These findings demonstrate that policymakers must stop relying on human oversight to protect against the harms of government algorithms. Policymakers must develop an alternative approach that more rigorously and democratically protects against harm.

### 5.1. The upper bound of human oversight

Before describing an alternative to human oversight policies, it is important to consider the upper bound of human oversight policies. To what extent can improved human oversight alleviate the flaws described above?

This is an open empirical question: some improvement is likely possible through sustained research in computer science and related fields. Such work could discover and develop interventions that improve the quality of human oversight and human-algorithm collaborations. In order to achieve this goal, it is necessary to take an "algorithm-in-the-loop" approach that focuses on improving human decision-making rather than optimizing algorithm performance (Green & Chen, 2019a). Notably, recent work along these lines has found that algorithmic accuracy does not always lead to the optimal outcomes (Bansal, Nushi, et al., 2021; Elmalech et al., 2015; Green & Chen, 2021; McCradden, 2021). It is therefore necessary to explore novel approaches for integrating algorithms into human decision-making processes. Promising directions worthy of further evaluation include providing decision support tools instead of specific recommendations (Yang et al., 2019) and adding greater structure to human-algorithm collaborations (Strandburg, 2021). For instance, cognitive forcing functions (e.g., prompting people to make a preliminary decision before being shown an algorithm's suggestion) can improve human-algorithm collaborations (Buçinca et al., 2021; Green & Chen, 2019b). Future research should also evaluate mechanisms that have been implemented in practice, such as providing training to human operators (Allegheny County Department of Human Services, 2019a) and requiring written justification and supervisor approval for any overrides (Allegheny County Department of Human Services, 2019a; Steinhart, 2006).

However, there are several reasons to be skeptical that the flaws of human oversight could be fully remedied through training and design mechanisms. First, sociotechnical collaborations are notoriously challenging. While it is common to blame "operator error" following harmful technological incidents, these incidents are typically caused by the structure of sociotechnical systems (Perrow, 1999). One of the central "ironies of automation" is that asking people to oversee automated systems creates "an impossible task" for the human overseer (Bainbridge, 1983). The underlying problem is a mismatch of skills and responsibilities: automated systems are typically adopted because they outperform human operators (at least along certain dimensions), yet then those same human operators are tasked with monitoring the automated systems. Human overseers are left with a task that is more difficult than their original charge. In the case of human oversight, algorithms are adopted because they make more accurate predictions than people do. Human oversight therefore means asking people to perform quality control for systems that perform at a higher prediction quality than people do, and often in inscrutable ways. Evaluating the quality of an algorithmic prediction is more difficult than simply making a prediction on one's own. Even greater training cannot overcome the challenges raised by human oversight of technology. For instance, although training can alleviate some aspects of automation bias, training cannot eliminate automation bias (Parasuraman & Manzey, 2010; Skitka et al., 1999).

Second, human oversight policies lack a clear goal or measure of success, as they are grounded in contradictory and circular logic. The respective motivations for algorithmic and human judgments are directly opposed. On the one hand, algorithmic decision-making is attractive because it promises consistency and rule-following. On the other hand, human oversight is attractive because it promises flexibility and discretion. Human oversight policies call for both rules and discretion without acknowledging the inherent tension between these goals. As a result, human oversight policies yield impossible-to-satisfy guidance. The Wisconsin Supreme Court's reasoning in *State v. Loomis* provides the most notable example of this circularity: the Court argued that judges should use risk assessments, but only on the condition that the algorithms do not actually alter decisions. In its attempt to attain both rule-following and discretion, the Court walked itself into a contradiction. If a risk assessment is to improve decisions, then it must alter some decisions. If a risk assessment cannot alter any decisions, then there is no reason to use it. Facilitating desirable combinations of human and algorithmic decision-making will ultimately require grappling with the inherent tensions between rules and discretion and considering the appropriate role for each in light of the trade-offs.

Third, even if people could provide effective oversight of algorithms, human oversight policies would still legitimize unjust algorithms and diminish accountability for the agencies and vendors who deploy these algorithms. In its ideal form, human overseers would follow accurate algorithmic judgments and override inaccurate or biased algorithmic judgments. By providing effective quality control over algorithmic recommendations, this idealized human oversight would prevent the harms that arise from algorithmic inaccuracy and bias. However, even this idealized human oversight cannot prevent the harms of algorithms that violate human rights, expand surveillance, or entrench inequity. For instance, critics of facial recognition, pretrial risk assessments, and predictive policing emphasize the fundamental injustices that would persist even (or especially) if these tools operated with greater accuracy (European Digital Rights, 2021a; Green, 2021; Hartzog & Selinger, 2018; Richardson et al., 2019; Stark, 2019; Stop LAPD Spying Coalition, 2018). In fact, more effective human oversight would further legitimize an agency's decision to adopt these systems. With a more effective form of quality control in place, it would become harder for critics to argue that these algorithms should not be deployed at all and to hold institutional leaders accountable for the harms of these algorithms.

## 5.2. *Institutional approach for overseeing government algorithms*

If legislators cannot depend on human oversight, then how should they regulate government algorithms? The answer is certainly not to simply remove human oversight, allowing algorithms to operate autonomously. However, it is also imprudent to entirely reject algorithms, relying solely on human judgment for all decisions. Instead, learning from the two flaws of human oversight policies, it is necessary to develop an alternative strategy for determining whether (and in what form) to incorporate algorithms into government decision-making.

In this section, I propose an institutional oversight approach to governing public sector algorithms. The central principle guiding this proposal is to promote greater rigor and democratic participation in government decisions about whether and how to use algorithms. While human oversight policies allow agencies to adopt algorithms as long as there is human oversight, this proposal increases the burden that agencies must overcome before they are permitted to implement an algorithm in practice (at least above a certain threshold of high-stakes applications). In doing so, this approach shifts accountability for algorithmic harms from human overseers to institutional leaders.

This approach operates in two stages. In the first stage, agencies must produce a written report justifying that it is appropriate to incorporate the algorithm into decision-making and that any proposed forms of human oversight are supported by empirical evidence. In the second stage, these reports must made public and approved through a participatory review process.

### 5.2.1. *Stage 1: Agency justification and evaluation*

Before government agencies incorporate algorithms into decision-making procedures, they must affirmatively justify that it is appropriate to integrate a proposed algorithm into the given decision-making process. In other words, rather than rely on human oversight to provide cover for the significant risks of algorithms, agencies must demonstrate that it is appropriate to use an algorithm at all.[18] This justification requires answering two central questions.

*5.2.1.1. Question 1: Is it appropriate to incorporate the algorithm into decision-making?* Determining whether it is appropriate to incorporate an algorithm into decision-making requires evaluating three factors. Policymakers must first consider "red lines" that mark unacceptable uses of algorithms. Many academics and civil society organizations across Europe and North America have argued that red lines are necessary for regulating AI (European Digital Rights, 2021a). For instance, scholars and communities argue that applications such as facial recognition and predictive policing violate fundamental notions of justice and human rights (Hartzog & Selinger, 2018; Richardson et al., 2019; Stark, 2019; Stop LAPD Spying Coalition, 2018). The issues with these algorithms cannot be remedied by increased accuracy or more reliable human oversight. Instead, it necessary to ban these applications of algorithms

---

[18] It is important to note that not adopting an algorithm does not require doing nothing and leaving the status quo in place. There are many potential avenues for reform and roles for algorithms beyond those typically proposed by vendors and agencies (Green, 2021). Indeed, as a broader political matter, agencies and the public should also consider whether using a particular algorithm actually represents a desirable approach to reform. This process would ideally include public participation at the upstream stage of formulating what applications of algorithms are worth exploring in the first place.

| Table 1 – The appropriate roles for human and algorithmic decision-making, based on the need for discretion within the given decision and the algorithm's trustworthiness.[20] | | | |
| --- | --- | --- | --- |
| | | Need for Discretion | |
| | | Low | High |
| (Relative) Trustworthiness of Algorithm | Low | 1) Primarily or solely human decision-making, with algorithms involved to the extent that rigorous research demonstrates benefits. | 2) Solely human decision-making. |
| | High | 3) Primarily or solely algorithmic decision-making. | 4) Primarily or solely human decision-making, with algorithms involved to the extent that rigorous research demonstrates benefits. |

outright. In recognition of these dangers, numerous jurisdictions across the US have recently passed bans and moratoria on police use of facial recognition and predictive policing algorithms (Hill, 2021; Ibarra, 2020; Stein, 2020).

If an algorithm is not prohibited by a red line, policymakers should then consider whether it can be appropriately integrated into a given decision. Making this determination involves two dimensions of analysis: one focused on the decision and one focused on the algorithm. The first dimension of analysis is the extent to which the decision in question is amenable to algorithmic decision-making. As with decision-support systems in other domains (Cummings, 2006), the appropriate role for algorithms depends on the extent to which human discretion is essential to making the decision. Because algorithms make decisions according to predetermined rules, the more that a decision requires individualized human discretion, the less appropriate it is for algorithms to play a role in decision-making. While it may be appropriate to automate decisions guided by predetermined rules, decisions guided by standards require human discretion and cannot be adequately made by algorithms (Citron, 2008). Discretion is particularly desirable for decisions that require determining the appropriate application of ambiguous and conflicting goals in individual cases that are difficult to classify in advance (Binns, 2020; Zacka, 2017). This analysis suggests that it may be appropriate for governments to adopt machine learning algorithms for pure prediction problems, but not for decisions that involve balancing predictions with other factors.

The second dimension of analysis is the extent to which the algorithm in question is trustworthy, relative to the stakes of the decision at hand. By trustworthy, I refer to an institutional analysis (does the algorithm provide validated and reliable advice?) rather than an individual analysis (do people trust the algorithm?). This distinction is particularly important given evidence that human decision-makers often mistrust accurate algorithmic predictions and place undue trust in flawed algorithms (Bansal, Wu, et al., 2021; Green & Chen, 2019a, 2019b; Jacobs et al., 2021; Lai & Tan, 2019). Trustworthiness requires a broader, rigorous evaluation of an algorithm's quality. The more that an algorithm is trustworthy, the more appropriate it is for the algorithm to influence decisions.

The trustworthiness of an algorithm depends on several factors. First and foremost, the algorithm must be rigorously evaluated for the task at hand. For an algorithm to be considered trustworthy, these tests should demonstrate that the algorithm makes predictions accurately and fairly. These evaluations should also demonstrate that the outcome of interest can be measured with reasonable accuracy and validity, as low-quality data and bad proxy variables present a significant limit on an algorithm's reliability (Jacobs & Wallach, 2021). Second, the algorithm must be transparent. Transparency into the algorithm's source code, training data, and development process are necessary for understanding how the algorithm works and how to interpret evaluations of the algorithm. Finally, trustworthiness is relative to the stakes of the decision: decisions that involve higher stakes associated with erroneous predictions require a higher standard for validating algorithms. These principles suggest that it may be appropriate for governments to adopt algorithms which have passed rigorous and transparent evaluations, but not algorithms which have known flaws or are hidden from public scrutiny.

Considering both discretion and trustworthiness in tandem can inform the appropriate role for an algorithm within a particular decision. Each dimension yields a general principle regarding algorithmic decision-making. The more that the decision requires discretion, the less appropriate it is to incorporate an algorithm into that decision. Similarly, the less that the algorithm is trustworthy, the less appropriate it is to incorporate that algorithm into a decision.

These two principles combine, as summarized in Table 1. In quadrants 2 and 3, these principles suggest relatively single-mode decision-making processes. If there is a high need for human discretion and an untrustworthy algorithm (quadrant 2), then governments should rely solely on human decision-making.[19] Conversely, if there is a low need for human discretion and a trustworthy algorithm (quadrant 3), then governments should rely primarily on algorithmic decision-making. In quadrant 3, it may be appropriate to incorporate human oversight in a relatively supervisory role, but too much human involvement could in fact diminish the quality of decisions.

---

[19] Relying on solely human decision-making does not mean relying on unchecked discretion. There are many existing mechanisms that constrain—without eliminating—the discretion of street-level bureaucrats (Lipsky, 2010).

[20] Of course, many scenarios will not fit neatly into this 2x2 grid. Table 1 can therefore be seen as identifying poles of a

In quadrants 1 and 4, these principles suggest potentially hybrid decision-making processes. If there is an untrustworthy algorithm and a low need for discretion (quadrant 1), or a trustworthy algorithm and a high need for discretion (quadrant 4), the principles for human versus algorithmic decision-making conflict with one another. Although governments should not turn to solely algorithmic decision-making in these scenarios, it is possible that incorporating the algorithm into human decision-making could improve outcomes. The default in both of these scenarios should be to retain solely human decision-making. Judgments about whether (and in what form) to incorporate an algorithm should depend on case-by-case empirical evaluations of whether the algorithm improves human decision-making.[21]

In sum, quadrants 1, 3, and 4 all involve a presumed or potential role for algorithms in decision-making. These scenarios raise the question of whether and how to combine human and algorithmic judgments in practice. In order to make these decisions, policymakers must turn to empirical evaluations regarding how people interact with the algorithm in the given setting.

*5.2.1.2. Question 2: How should the algorithm be integrated with human decision-making?* When analysis of Question 1 suggests that there might be a role for algorithms in collaboration with humans (i.e., quadrants 1, 3, and 4 in Table 1), policymakers must then determine whether and in what form to combine human and algorithmic judgments. Unless an algorithm is intended to operate autonomously, it is not sufficient to show that the algorithm is trustworthy on its own. Instead, there must be evidence suggesting that people can oversee the algorithm and that incorporating the algorithm into decision-making will improve outcomes. These evaluations must precede any decision to adopt algorithms in a manner that involves human-algorithm collaborations.

The primary method for assessing human oversight is to conduct experimental evaluations of human-algorithm collaborations before implementing an algorithm in practice. First, in order to uncover breakdowns in human-algorithm collaboration and to experiment with potential remedies, developers should study how laypeople use the algorithm in a lab setting. Although these experiments would be with laypeople, they can shed light on some behaviors of experts in practice and can be conducted in a quick, low-stakes manner on platforms such as Amazon Mechanical Turk (Green & Chen, 2021). These initial experiments would provide a baseline of evidence regarding human-algorithm collaborations and suggest strategies for improving these collaborations. Second, once preliminary evidence suggests that people can collaborate effectively with the algorithm, developers and governments should study how practitioners use the algorithm in a lab setting. These experiments can test the mechanisms identified as most effective with laypeople and determine the likely effects of implementing the algorithm.[22] An algorithm should be incorporated into practice only when these preliminary evaluations suggest that adopting the algorithm will improve human decision-making and that people are able to perform the desired oversight functions.

Finally, preliminary evaluations of human-algorithm collaborations should be supplemented with an agency's plan for how it will monitor human oversight even after an algorithm is adopted in practice. Persistent monitoring is particularly important in light of evidence that judicial uses of algorithms can shift over time (Stevenson, 2018) and that practitioner responses to algorithms depend on localized details of institutional implementation (Brayne & Christin, 2020). Agencies should be required to collect information about human interactions with algorithms, for instance by tracking overrides of algorithmic decisions to check for racial disparities (Steinhart, 2006). Furthermore, given that automation can reduce its users' sense of control, responsibility, and moral agency (Berberian et al., 2012; Cummings, 2006), it is important to continuously monitor whether the algorithm distorts or erodes the moral agency of decision-makers. Evaluations should ensure that algorithms complement rather than diminish the work of government staff (Pasquale, 2020).

### 5.2.2. Stage 2: Democratic review and approval

After an agency has provided a report justifying its intent to adopt an algorithm, this report must be subject to review and approval by the public or a democratically accountable body. This stage is essential to enabling democratic accountability regarding decisions to adopt algorithms. Although fostering robust public participation in algorithmic governance remains an open challenge, there are several existing processes that can be adapted for this purpose.

The appropriate form of public review will depend on the jurisdiction. At the local level, city departments could follow a procedure modeled on surveillance oversight ordinances in US municipalities. These policies enact a process for democratic review of surveillance technologies. For instance, per the Surveillance Technology Ordinance passed in Cambridge, Massachusetts in 2018, municipal departments must seek approval from the City Council before using surveillance technologies (City of Cambridge, 2018). Departments must first provide the City Council with reports describing the proposed uses, impacts, and governance of the technology. These reports must be made publicly available and discussed at a City Council meeting open to the public. Following this discussion, the City Council decides whether to approve or reject the department's proposal (City of Cambridge, 2018).

State and federal agencies could follow a process akin to the notice-and-comment procedure laid out by the US Ad-

---

two-dimensional spectrum. For instance, as we move from low-discretion to high-discretion decisions, the trustworthiness required to consider using an algorithm increases accordingly.

[21] The precise meaning of "improves" will depend on the context and goals of any specific decision-making process. Common standards of improvement include the accuracy of predictive judgments and the fairness of decisions.

[22] Strictly speaking, it is possible to run experiments only with practitioners. However, it will generally be beneficial to begin with experiments with laypeople. Compared to studies with experts, studies with laypeople can be conducted more quickly and with more participants, enabling deeper scientific inquiry (Green & Chen, 2021). As we gain a more thorough understanding of human-algorithm collaborations, experiments with laypeople may become less essential.

ministrative Procedure Act. Notice-and-comment provides a mechanism for transparency and public review of administrative rulemaking. When agencies want to create or modify an administrative regulation (i.e., rule), they must notify the public of their proposal, accept public comments, incorporate public comments into their final rule, and justify how the final rule aligns with those comments (Kerwin & Furlong, 2019). Because public participation is essential to the democratic legitimacy of this process, courts often review final rules in light of whether they were responsive to public comments (Kerwin & Furlong, 2019). Notice-and-comment should be extended to cover agency uses of algorithms, particularly given the errors and harms that have resulted from government algorithms that eschewed this process (Calo & Citron, 2021; Citron, 2008).

### 5.3.  *Benefits of institutional oversight approach*

This proposed institutional oversight process expands the scope of recent regulatory efforts to require proactive assessments of an algorithms. Already, several human oversight polices mandate that agencies or vendors must conduct proactive impact assessments that evaluate algorithms for accuracy and fairness (Brown, 2020; California Legislature, 2021; European Commission, 2021; Government of Canada, 2021; Washington State Legislature, 2020). Some policies also call for ongoing evaluations after implementation (Brown, 2020; European Commission, 2021). These requirements align closely with the principle of trustworthiness. Indeed, the EU AI Act explicitly describes its proactive assessment as intended "to ensure a high level of trustworthiness of high-risk AI systems" (European Commission, 2021).

However, existing regulations do not capture the components other than trustworthiness from the institutional oversight approach described above, in large part due to assumptions about the efficacy of human oversight. First, these policies rarely include red lines, and exceptions such as the EU AI Act include only a minimal set (European Commission, 2021; European Digital Rights, 2021b). Notably, the EU's High-Level Expert Group on AI did not include red lines (Metzinger, 2019), instead describing human oversight as the essential principle to ensure fundamental rights and human autonomy (High-Level Expert Group on AI, 2019). Second, although some scholars have suggested that decisions about whether to adopt an algorithm should be based on the nature of a decision-making task (Citron, 2008; Cummings, 2006), existing policies do not incorporate this consideration. This omission is likely due (at least in part) to the assumption that human oversight ensures that human discretion remains intact even when an algorithm is used. Third, no policies require predeployment evaluations of human-algorithm collaborations. As described throughout this paper, policies take for granted that human oversight is effective, without recourse to empirical evidence. Finally, current policies lack requirements that decisions to use algorithms undergo public input and approval processes.

The proposed institutional oversight approach would thus yield several improvements over the status quo of human oversight policies. First, and most directly, this approach would increase the rigor in any uses of algorithms that involve human oversight. Rather than take for granted that people can effectively oversee algorithms, policymakers must empirically evaluate whether any proposed forms of human oversight are actually effective. Given the empirical evidence demonstrating the limits of human oversight, the default assumption should be that human oversight is likely to be ineffective, unless proven otherwise. The burden should therefore fall on agencies proposing human oversight of algorithms to provide affirmative evidence that this mechanism actually improves outcomes and addresses concerns about algorithmic decision-making. This requirement remedies the first flaw of human oversight policies, in which human oversight is presented as a safeguard against algorithmic harms despite minimal empirical evidence that it actually provides reliable protections.

Second, the institutional oversight approach would promote a more stringent standard for decisions about whether and how to integrate algorithms into government decision-making. Rather than craft blanket rules that enable governments to use algorithms as long as a human provides oversight, policymakers must place greater scrutiny on whether an algorithm is even appropriate to use in a given context. This requirement remedies the second flaw of human oversight policies, which is that they legitimize the use of algorithms despite flaws suggesting that these tools should not be used at all.

Furthermore, requiring agencies to justify their decisions to adopt algorithms would help to ensure that institutional leaders are held accountable for algorithmic harms. Whereas human oversight policies direct attention to human operators, compelling agencies to justify their decisions to adopt algorithms would direct attention to vendors and agency leaders. This requirement would make it more difficult for vendors and institutional leaders to blame human operators when an algorithm produces harms.[23] The terms of debate after algorithmic harms arise would move upstream, from whether human operators exercised appropriate oversight to whether an algorithm should have been adopted in the first place.

Finally, the second stage of the institutional oversight approach would empower democratic participation in decisions regarding whether to adopt algorithms. The legislative developments of recent years demonstrate that if the public is not granted sufficient authority to shape how governments use algorithms, agencies and companies will often ignore public concerns. In Europe, for instance, communities and companies have jockeyed over the proper role for red lines. As the European Commission was developing its AI Act proposal, a coalition of 62 civil society organizations preemptively called for red lines (European Digital Rights, 2021a).[24] However, al-though the Act prohibits some uses of AI (such as social scoring by public authorities), it authorizes most of the applications highlighted by this coalition, merely labeling such applications as "high-risk" (European Digital Rights, 2021b). Following critiques by civil society organizations, legislators responsible for amending the Act have called for stronger bans on facial recognition, setting up future showdowns with member states and technology companies that support facial recognition (Kelly, 2021). Meanwhile, successful efforts to curb government algorithms in the United States—such as recent facial recognition and predictive policing bans—have required concerted local advocacy that overcame significant resistance from agencies (Hill, 2021; Miller, 2020).

---

[23] This should apply even in many cases in which human operators may appear to be at fault for failing to provide proper oversight. If agency leaders choose to implement a flawed or unjust algorithm without any evidence that staff can provide the desired form of oversight, then they should not be permitted to foist blame on staff for failing to do a task that never should have been expected of them. Human operators should not be completely immune from accountability, but their accountability should be tempered in light of whether the algorithm in question is appropriate for the agency to be using at all and whether people can be reasonably expected to provide the desired forms of oversight.

[24] The coalition's open letter highlighted five categories of AI particularly deserving of red lines: biometric mass surveillance,

## 6. Conclusion

This study evaluated the global policy trend toward requiring human oversight of algorithms used by governments. By considering human oversight policies in light of research on human-algorithm interactions, I found that these policies suffer from two significant flaws. First, the vast majority of evidence suggests that people cannot adequately provide the envisioned forms of oversight. Second, the incorrect assumption of effective human oversight legitimizes the use of flawed and unaccountable algorithms in government. Thus, rather than enable governments to attain the benefits of algorithms without incurring the associated risks, human oversight policies justify inappropriate uses of algorithms and hinder accountability for institutional decision-makers.

In light of these findings, I proposed a shift from human oversight to institutional oversight as the primary mechanism for promoting the quality and legitimacy of algorithm-assisted decisions. This institutional oversight process increases the burden on agencies to justify their decisions to adopt algorithms, helping to ensure that human oversight no longer operates as a superficial salve for fundamental concerns about algorithmic decision-making in government. My proposed institutional oversight approach involves two stages that precede an agency's ability to adopt an algorithm. First, agencies must justify that it is appropriate to adopt an algorithm and that any proposed forms of human oversight are supported by empirical evidence. Second, agencies must make these written justifications publicly available and receive approval through a democratic review process.

As governments adopt algorithms to make or inform consequential decisions, regulation is necessary to avoid producing injustices and violating fundamental legal principles. It is not enough merely to enact regulations, however: policymak-ers must ensure that their regulations actually provide the desired protections and benefits. Relying on intuitively appealing but ineffective regulation could lead to the worst of both worlds: the underlying problem persists, yet the presence of the regulation leads to the perception that the problem has been solved. Efforts to regulate government algorithms must therefore be particularly attentive to the social contexts in which algorithms are embedded and to empirical evidence about how algorithms are implemented. By taking a more sociotechnical and evidence-based regulatory approach, policymakers will facilitate more democratic and equitable decisions about algorithmic governance.

## Declaration of Competing Interest

The authors declare that they have no known competing financial interests or personal relationships that could have appeared to influence the work reported in this paper.

## Data Availability

No data was used for the research described in the article.

## Acknowledgments

Thank you to Amit Haim, Amba Kak, Ben Lempert, Meg Leta Jones, Nicholson Price, Andrew Schrock, Salomé Viljoen, and NYU's Privacy Research Group for insightful feedback on earlier drafts of this article. I am grateful to Alison Christiansen for providing excellent research assistance. I also thank the reviewers for their generous and helpful suggestions.

## Appendix: Summary of human oversight policies

This table summarizes the 41 policy documents that I reviewed as part of this study. Document Classification refers to the type of policy document (1 = proposed or passed legislation; 2 = policy guidance by government or government-appointed bodies; 3 = manuals, policies, and court cases related to the Allegheny County Family Screening Tool and to risk assessments used in US criminal justice settings). Approach to Human Oversight refers to how each policy presents the appropriate role for human oversight (1 = restricting solely automated decisions; 2 = emphasizing human discretion; 3 = requiring meaningful human input). Documents are grouped by their approach to human oversight.

---

border and migration control, social scoring systems, predictive policing, and risk assessments in the criminal justice system (European Digital Rights, 2021a).

| Author/Publisher | Title | Year | Document Classification | Approach to Human Oversight |
|---|---|---|---|---|
| Austrian Parliament | Datenschutzgesetz (Data Protection Act) | 2018 | 1 | 1 |
| Belgian Federal Parliament | Loi relative à la protection des personnes physiques à l'égard des traitements de données à caractère personnel (Law on the Protection of Natural Persons with regard to the Processing of Personal Data) | 2018 | 1 | 1 |
| Bundestag (German Parliament) | Federal Data Protection Act (BDSG) | 2019 | 1 | 1 |
| Dutch Parliament | Uitvoeringswet Algemene gegevensbescherming (Implementation Act General Data Protection Regulation) | 2018 | 1 | 1 |
| European Parliament and the Council of the European Union | General Data Protection Regulation (GDPR) | 2016 | 1 | 1 |
| European Parliament and the Council of the European Union | Data Protection Law Enforcement Directive | 2016 | 1 | 1 |
| French Parliament | French Data Protection Act | 2018 | 1 | 1 |
| Houses of the Oireachtas (Irish Parliament) | Data Protection Act 2018 | 2018 | 1 | 1 |
| Hungarian Parliament | Data Protection Act | 2018 | 1 | 1 |
| Kingdom of Bahrain | Personal Data Protection Law | 2018 | 1 | 1 |
| National Assembly of Québec | Bill 64: An Act to modernize legislative provisions as regards the protection of personal information | 2020 | 1 | 1 |
| National Congress of Brazil | General Data Protection Law | 2019 | 1 | 1 |
| Parliament of Mauritius | The Data Protection Act 2017 | 2017 | 1 | 1 |
| Republic of Argentina | Ley De Protección De Los Datos Personales (Personal Data Protection Law) | 2018 | 1 | 1 |
| Republic of Kenya | The Data Protection Act, 2019 | 2019 | 1 | 1 |
| Republic of South Africa | Protection of Personal Information Act | 2013 | 1 | 1 |
| Republic of Uganda | The Data Protection and Privacy Act, 2019 | 2019 | 1 | 1 |
| Senator Sherrod Brown | Data Accountability and Transparency Act of 2020 | 2020 | 1 | 1 |
| Slovenian Parliament | Zakon o varstvu osebnih podatkov na področju obravnavanja kaznivih dejanj (Personal Data Protection Act in the field of dealing with criminal offenses) | 2020 | 1 | 1 |
| UK Parliament | Data Protection Act 2018 | 2018 | 1 | 1 |
| Allegheny County Department of Human Services | Frequently-Asked Questions | 2019 | 3 | 2 |
| Allegheny County Department of Human Services | Ethical Analysis: Predictive Risk Models at Call Screening for Allegheny County | 2019 | 3 | 2 |
| Annie E. Casey Foundation | Juvenile Detention Risk Assessment: A Practice Guide for Juvenile Detention Reform | 2006 | 3 | 2 |
| Arnold Ventures | Public Safety Assessment FAQs ("PSA 101") | 2019 | 3 | 2 |
| Australian Human Rights Commission | Human Rights and Technology: Discussion Paper | 2019 | 2 | 2 |
| Government of Canada | Directive on Automated Decision-Making | 2021 | 1 | 2 |
| Indiana Supreme Court | Malenchik v. State | 2010 | 3 | 2 |
| Multi-Health Systems Inc. | The Level of Service Inventory-Revised Manual | 2001 | 3 | 2 |
| New Jersey Courts | One Year Criminal Justice Reform Report to the Governor and the Legislature | 2017 | 3 | 2 |
| Northpointe | Practitioner's Guide to COMPAS Core | 2015 | 3 | 2 |
| Statistics New Zealand | Algorithm Assessment Report | 2018 | 2 | 2 |
| Statistics New Zealand | Algorithm Charter for Aotearoa New Zealand | 2020 | 2 | 2 |
| Wisconsin Dept of Corrections | Electronic Case Reference Manual | 2018 | 3 | 2 |
| Wisconsin Supreme Court | Wisconsin v. Loomis | 2016 | 3 | 2 |
| Administrative Conference of the United States | Government by Algorithm: Artificial Intelligence in Federal Administrative Agencies | 2020 | 2 | 3 |

| Author/Publisher | Title | Year | Document Classification | Approach to Human Oversight |
|---|---|---|---|---|
| **Article 29 Data Protection Working Party** | Guidelines on Automated individual decision-making and Profiling for the purposes of Regulation 2016/679 | 2018 | 2 | 3 |
| **European Commission** | Proposal for a Regulation laying down harmonised rules on artificial intelligence (Artificial Intelligence Act) | 2021 | 1 | 3 |
| **High-Level Expert Group on AI** | Ethics Guidelines for Trustworthy AI | 2019 | 2 | 3 |
| **High-Level Expert Group on AI** | Assessment List for Trustworthy Artificial Intelligence | 2020 | 2 | 3 |
| **UK Information Commissioner's Office** | Guidance on the AI Auditing Framework | 2020 | 2 | 3 |
| **Washington State Legislature** | SB 6280 - 2019-20: Concerning the use of facial recognition services | 2020 | 1 | 3 |

## Author Information

Ben Green is a Postdoctoral Scholar in the Michigan Society of Fellows and an Assistant Professor in the Gerald R. Ford School of Public Policy. He holds a PhD in Applied Mathematics, with a secondary field in Science, Technology, and Society, from Harvard University. His book, *The Smart Enough City: Putting Technology in Its Place to Reclaim Our Urban Future*, was published in 2019 by MIT Press. Ben is also an Affiliate at the Berkman Klein Center for Internet & Society at Harvard and a Fellow at the Center for Democracy & Technology.